\documentclass{article}
\usepackage{hyperref}
\usepackage{amsmath}
\usepackage{amsfonts}
\usepackage[round]{natbib}
\newcommand{\ED}{\mathbb{E}\left(D\right)}

\title{\vspace{-2.0cm}The dual-Barab\'asi-Albert model}
\author{Niema Moshiri}
\date{}

\begin{document}

\maketitle

\begin{abstract}
The ability to sample random networks that can accurately represent real social contact networks is essential to the study of viral epidemics. The Barab\'asi-Albert model and its extensions attempt to capture reality by generating networks with power-law degree distributions, but properties of the resulting distributions (e.g. minimum, average, and maximum degree) are often unrealistic of the social contacts the models attempt to capture. I propose a novel extension of the Barab\'asi-Albert model, which I call the ``dual-Barab\'asi-Albert'' (DBA) model, that attempts to better capture these properties of real networks of social contact.
\end{abstract}

\section{Introduction}
Simulation has become a useful tool in the study of viral epidemics, and the ability to capture properties of real social contact networks (e.g. sexual networks) is essential to generating useful simulated datasets \citep{Moshiri2018}
The Barab\'asi-Albert (BA) model is widely-used when modeling large social networks \citep{Barabasi1999}.
The BA model is parameterized by $n$ and $m$ and produces a graph with $n$ vertices and $2m$ edges such that the degree distribution follows a power-law.
The BA model begins by initializing an empty graph with $m$ vertices and iteratively adds $n-m$ new vertices. For each new vertex that is added, $m$ new edges are also added. One end of the edge is attached to the new vertex, and the other is attached to an existing vertex via preferential selection.

Although the BA model yields networks with power-law degree distributions, they often have properties that are unrealistic of real sexual contact networks. For example, when modeling sexual contacts, one may want to have an expected degree of 8 sexual partners per individual, which would be achieved by simulating a BA graph with $m=4$. However, the resulting graph would have a minimum degree of 4, whereas true sexual contact networks typically have a large proportion of individuals who are monogamous (i.e., degree of 1). Under the BA model, a minimum degree of 1 can only be achieved by using $m=1$, but the expected degree would be 2. Extensions of the BA model exist that alleviate other potentially unrealistic properties of networks simulated under the BA model \citep{Albert2000}, but these issues in degree distribution remain.

I propose a novel extension of the Barab\'asi-Albert model, which I call the ``dual-Barab\'asi-Albert'' (DBA) model, to better capture these properties.

\section{Model Description}
The DBA model is parameterized by $n$, $m_1$, $m_2$, and $p$ such that $1\le m_1,m_2<n$ and $0\le p\le 1$. The DBA model begins by initializing an empty graph with $\max(m_1,m_2)$ vertices. Then, the remaining $n-\max(m_1,m_2)$ vertices are added iteratively. For each new vertex, with probability $p$, $m_1$ new edges are added, and with probability $1-p$, $m_2$ new edges are added. The new edges are added in the same manner as in the BA model \citep{Barabasi1999}.

\section{Theoretical Properties}
\subsection{Expected Degree}
By definition, the expected degree $\ED$ is the expected total degree divided by the total number of vertices. Also by definition, the expected total degree is exactly twice the expected total number of edges. For each of the new $n-\max(m_1,m_2)$ vertices, either $m_1$ or $m_2$ edges are added (with probability $p$ and $1-p$, respectively). Thus, the expected degree is the following:

\begin{equation}
\ED=2\left(n-\max(m_1,m_2)\right)\left(pm_1+(1-p)m_2\right)/n
\end{equation}

Asymptotically, $n>>m_1,m_2$, so the following expected degree is observed:

\begin{equation}
\lim_{n\to\infty}\ED=2\left(pm_1+(1-p)m_2\right)
\end{equation}

\subsection{Degree Distribution}
In the BA model with parameter $m$, the asymptotic degree distribution $p_k = \frac{2m(m+1)}{k(k+1)(k+2)}$ for $k\ge m$ \citep{Barabasi2016}. In the DBA model, the following is observed:

\begin{equation}
p_k=
\begin{cases}
\hfil 0 & \hfil k < \min(m_1,m_2)\\[10pt]
\hfil \frac{2pm_1(m_1+1)}{k(k+1)(k+2)} & \hfil \min(m_1,m_2) \leq k \leq \max(m_1,m_2)\\[10pt]
\hfil \frac{2\left(pm_1(m_1+1)+(1-p)m_2(m_2+1)\right)}{k(k+1)(k+2)} & \hfil \max(m_1,m_2) \leq k
\end{cases}
\end{equation}

\section{Discussion}
I proposed a novel extension of the Barab\'asi-Albert (BA) model called the ``dual-Barab\'asi-Albert`` (DBA) model, and I derived two useful properties of the model: the expected degree and the degree distribution of networks sampled under the model. However, these are only two properties of interest, and many useful statistical properties exist, such as average path length, clustering coefficient, etc.

It must be emphasized that, while one key feature of the BA model and many of its extensions is power-law degree distribution, the DBA model \textit{is not guaranteed} to (and will likely not) yield power-law degree distributions. On the contrary, without careful selection of $p$, the degree distribution will likely be bimodal.

The DBA model has been implemented in the \href{https://networkx.github.io/}{NetworkX} Python package.

\section*{Funding}
This work was supported by NIH subaward 5P30AI027767-28 to NM.

\bibliography{references}
\end{document}